%% file: main.tex
\documentclass[journal,transmag]{IEEEtran}

\usepackage{graphicx}
\usepackage[utf8]{inputenc}
\usepackage{pdfpages}
\usepackage{graphicx} 
\usepackage{color}  
\usepackage{mathtools} 
\usepackage{amsmath}
\usepackage{amsfonts}
\usepackage{amssymb}
\usepackage{enumerate}
\usepackage{booktabs}
\usepackage{lipsum}
\usepackage{float}
\usepackage[section]{placeins}
\usepackage{placeins}
\usepackage{todonotes}
\usepackage{empheq}
\usepackage{amsmath}
\usepackage{listings}
\usepackage{tcolorbox}
\usepackage{ragged2e}
\usepackage[linesnumbered]{algorithm2e}
\usepackage{amssymb}
\usepackage{lineno}

\begin{document}

\title{Autonomic Intrusion Response in \\ Distributed Computing using Big Data}

\author{
\IEEEauthorblockN{
Kleber Vieira\IEEEauthorrefmark{1},
Fernando Koch\IEEEauthorrefmark{1,2},
Jo\~{a}o Bosco Mangueira Sobral\IEEEauthorrefmark{1}, \\
Carlos Becker Westphall\IEEEauthorrefmark{1}, and
Jorge Lopes de Souza Leão\IEEEauthorrefmark{3}}

\IEEEauthorblockA{\IEEEauthorrefmark{1}Network and Management Laboratory (LRG), Federal University of Santa Catarina (UFSC), Brazil}\IEEEauthorblockA{\IEEEauthorrefmark{2}School of Computing and Information Systems, The University of Melbourne, Australia}
\IEEEauthorblockA{\IEEEauthorrefmark{3}Technology Centre, Federal University of Rio de Janeiro (UFRJ), Brazil}}

\IEEEtitleabstractindextext{
\begin{abstract}

\input{00abstract.tex}
\end{abstract}

\begin{IEEEkeywords}
Intrusion Detection Systems, Cybersecurity,  Distributed Computing, Big Data, Autonomic Computing
\end{IEEEkeywords}}

\maketitle
\IEEEdisplaynontitleabstractindextext
\IEEEpeerreviewmaketitle

\section{Introduction}
\label{sec-intro}
\input{01intro.tex}

\section{Background and Related Work}
\label{sec-background}
\input{02background.tex}

\section{Proposal}
\label{sec-proposal}
\input{03proposal.tex}

\section{Results}
\label{sec-results}
\input{04results.tex}

\section{Conclusions}
\label{sec-conclusion}
\input{05conclusion.tex}

\section*{Acknowledgement}
\input{06acknowledge.tex}


\bibliographystyle{ieeetr}
\bibliography{mybib}

\end{document}

%% file: 00abstract.tex

We introduce a method for Intrusion Detection based on the classification, understanding and prediction of behavioural deviance and potential threats, issuing recommendations, and acting to address eminent issues. Our work seeks a practical solutions to automate the process of identification and response to Cybersecurity threats in hybrid Distributed Computing environments through the analysis of large datasets generated during operations. We are motivated by the growth in utilisation of Cloud Computing and Edge Computing as the technology for business and social solutions. The technology mix and complex operation render these environments target to attacks like hijacking, man-in-the-middle, denial of service, phishing, and others. The  \emph{Autonomous Intrusion Response System} implements innovative models of data analysis and context-aware recommendation systems to respond to attacks and self-healing. We introduce a proof-of-concept implementation and evaluate against datasets from experimentation scenarios based on public and private clouds. The results present significant improvement in response effectiveness and potential to scale to large environments.

%% file: 01intro.tex


There is a growing number of cybersecurity threats related to the extended utilisation of Cloud Computing and Edge Computing. The \emph{Brazilian Center for Studies, Response and Treatment of Security Incidents (CERT.br)}, which monitors attacks attempts and their types, shows the growing tendency of Cyberattack incidents such as Distributed Denial of Service (DDoS) attacks \cite{lau2000distributed,chang2002defending} whose incidents grew by 125.36\% between the first quarter of 2016 with the same period of 2015. Such attempts, successful or not, result in economic, reputation, and social impact. A report from PwC Consulting describes the economic impact of Cybersecurity breaches in areas like disruption of operations and manufacturing, compromise of sensitive data, negative impact to product and services, damage of physical property, and harm to human life \cite{pwc2018}. The scaling number of virtual crimes and the exploitation of vulnerabilities in Distributed Computing demand new forms of preventive measures to preserve security and privacy. 


We are motivated by the need for effective Cybersecurity strategies for intrusion detection and fast response, aiming to prevent disruption, preserve privacy and security, and optimise operations. Cybersecurity threats are primarily linked to storage and transfer of large chunks of information, along with their importance and vulnerability \cite{smith2002issues}. The most common security threats in Distributed Computing include hijacking, man-in-the-middle, denial of service, phishing, and others \cite{pfleeger2002security, subashini2011survey, behl2011emerging}. Buyya \emph{et al} \cite{Buyya2012} points the lack of a well-defined security strategies in heterogeneous environments combining Cloud Computing and Edge Computing. This issues is mostly due to the characteristics of the environment involving distributed architectures, complex and heterogeneous elements, and large scale operations.


We are looking into a combination of \emph{Autonomic Computing} \cite{kephart2003vision, horn2001autonomic} and \emph{Big Data} to deal with the large volume of information collected from the audits of the various system components and to provide rapid response. This work contributes to the state-of-the-art by:

\begin{itemize}
\item Providing a reference architecture for \emph{Autonomic Intrusion Response System} based on a combination between Autonomic Systems and Big Data.
\item Presenting a proof-of-concept implementation of a full-cycle attack-response interaction in heterogeneous Distributed Computing environments.
\item Analysing this approach's performance for accuracy, efficiency, and scalability to real-world scenarios.
\end{itemize}

In what follows, we elaborate on the background, state-of-the-art, and technology gap. Section \ref{sec-proposal} outlines our proposal. Section \ref{sec-results} describes the results from executing a proof-of-concept implementation upon experimentation environments of private and public clouds. We discuss the results and opportunities in Section \ref{sec-conclusion}.

%% file: 02background.tex

System administrators demand approaches of Intrusion Detection Systems (IDS) to minimise the harms of hackers, crackers, and other cyber-criminals \cite{modi2013survey,schulter2006towards,dali2015}. In general, preventive systems employ techniques to analyse the behaviour and origin of the attempts to then define whether the action is allowed \cite{stakhanova2007taxonomy}. Response time is crucial to prevent intrusions. Cohen \emph{et al} \cite{cohen1999simulating} points out that for a skillful intruder his attack will have 80\% chance of success if the response time is around 10 hours, 95\% chance if the intruder has 20 hours, and for over 30 hours the attack renders virtually infallible; however, if the response is immediate, then the chances of the intruder's success are practically nil. Nonetheless, current approaches present a significant time gap between detection and response, mostly due to the need for manual intervention \cite{northcutt2002network,carver2000intrusion, kholidy2016risk}.

There is a cohort of research looking into how to improve IDS towards quick detection of malicious or unauthorised actions \cite{debar2000revised,kumar2015survey}, and intelligent management methods of Distribute Computing \cite{subashini2011survey, assuncao2004grids}. In general, an IDS encompasses:

\begin{enumerate}
\item \emph{Detection}, usually performed automatically by monitoring patterns in the systems' log entries and behaviour of the elements.

\item \emph{Warning}, triggered via analysis of behaviour patterns and raising awareness of potential issues to system administrators.

\item \emph{Decision making}, provides decision making support based on data analytics systems.

\item \emph{Response}, implementing  actions upon the elements, along with evaluation of the their results.
\end{enumerate}

Buyya \emph{et al} \cite{Buyya2012} argues that existing strategies for attack detection and response still fail to provide satisfactory results for Distributed Computing environments. Notably, current implementations presents a delay between \emph{Detection} and \emph{Response}. Moreover, current developments tend to focus on \emph{Detection} and \emph{Warning}, whereas there is a critical demand to optimise the manual intervention in \emph{Decision Making} and \emph{Response} \cite{stakhanova2007taxonomy}.

Hence, there is a technology gap between strategies for detecting attack attempts and existing response mechanisms. Although fast response is a clear demand, current implementations mostly depend on manual interventions rending these solutions slow and ineffective. In this context, the motivation for this work encompasses:
\begin{itemize}
\item define the requirements for an effective autonomic attack response;
\item outline the required decision-making algorithms to support this systems;
\item field test proposed approaches in controlled environments in order to evaluate performance, applicability to real-world scenarios, and scaleability to large Distribute Computing systems. 
\end{itemize}

%% file: 03proposal.tex

The \emph{Autonomic Intrusion Response System} (SARI) follows the vision of \emph{autonomic computing} around self-healing, self-protection and self-optimising. The solution works based on the Monitor-Analyse-Plan-Execute-Knowledge (MAP-K) architecture to efficiently analyse large amounts of data about the utilisation of Distribute Computing resources.

\begin{figure}
\centering
\includegraphics[width=\linewidth]{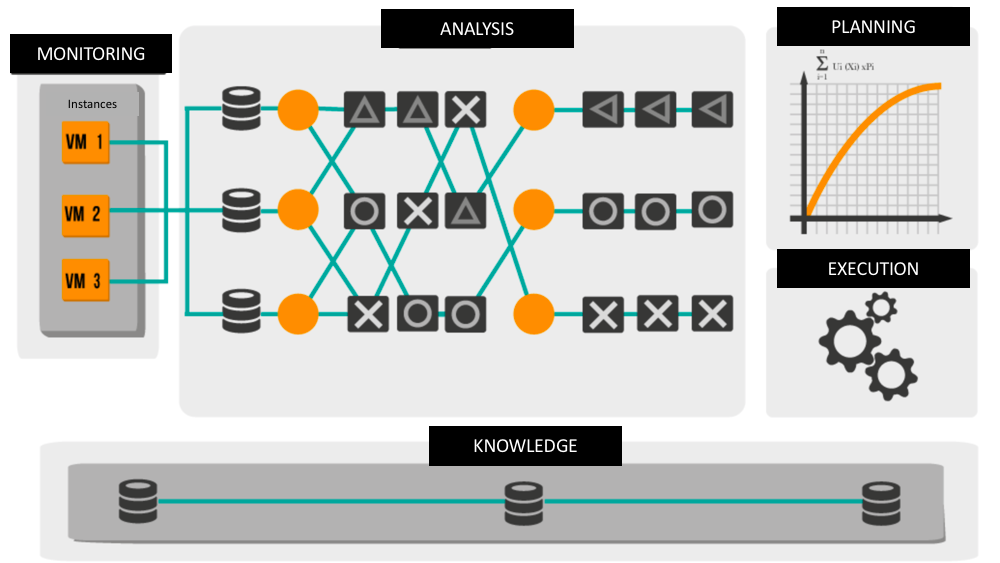}
\caption{Architecture of the \emph{Autonomic Intrusion Response System}}
\label{fig:sari}
\end{figure}

Figure \ref{fig:sari} depicts the system architecture. We devised an approach to collect \emph{system log} datasets about network traffic, system information, and sensors, following the proposal by Suthaharan \emph{et al} \cite{suthaharan2013big}. The solution pre-processes these datasets to consolidate information and remove \emph{noise}. The cycles for analysis and planning implement the \emph{MapReduce} strategy to correlate the information -- this is a programming model designed to process large volumes of data in parallel, dividing the work into a set of independent tasks \cite{dean2008mapreduce,ahn2014big}. Our architecture encompasses the following components:

\begin{itemize}

\item \emph{Monitoring Module} implements \emph{probes} to collect information about behaviour changes of the managed elements, and other execution information; these datasets include e.g. \emph{system log} and other monitoring systems installed in the Virtual Machines such as Snort, OSSEC, Hypervisor, network traffic, system settings, and SMNP data \cite{modi2013survey, werner2017cloud};  sensors are designed to collect data from \emph{Hypervisors} and VM instances through the library \emph{jNetPcap}.

\item \emph{Analysis Module} implements the processes for \emph{categorisation} (or mapping)  and \emph{reduction}; in this module, \emph{MapReduce} is applied to (i) identify the signatures of known attacks and (ii) extract significant data such as the origin of the attack, features of the data packages, and others. This process analyses and classifies data packages in relation to their protocol. Then, the process applies different algorithms to specific protocols that implement the process to reduce data volumes. The solution results in a data hierarchy  for analysis and a compilation of possible issues causing the attacks. 

\item \emph{Planning Module} implements the MAPE-K loop strategy based on the theory of expected utility. The planning component collects data from the \emph{Analysis module} to characterise the current situation. Then, the planning process applies algorithms to select the response action that is most likely to work in the determined scenario. The process applies the \emph{theory of expected utility} is applied to select the best response. The  to the attack. The method works by analysing diverse alternatives applicable to the situation  possible In this technique, the various possible alternatives are analysed and the one that brings the highest response value to a given environment configuration is selected \cite{bordley2009decision}.

\item \emph{Execution Module} performs the notification or response action on detected intrusions depending on the configuration.

\item \emph{Knowledge Module} holds the information requires for system's operations, such as: collected data; known signatures; time values; cost and probability of each response; applied response techniques; environment settings, and more.

\end{itemize}

\begin{figure*}
\centering
\includegraphics[width=\linewidth]{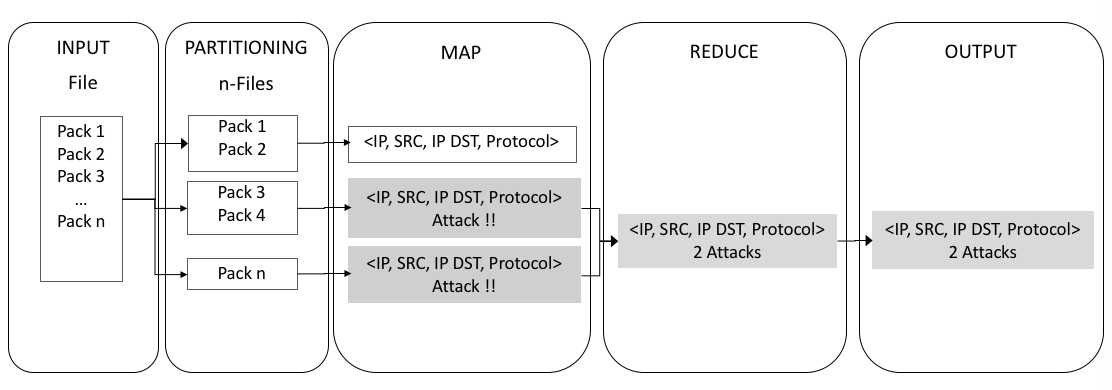}
\caption{Analysis module with MapReduce}
\label{fig:sari_map_reduce}
\end{figure*}

The solution employs a knowledge based approach to detect known attacks by comparing attack signatures to suspicious actions  \cite{vieira2010intrusion,vieira2015}. Figure \ref{fig:sari_map_reduce} depicts the operation. The strategy applies \emph{MapReduce} to allow working on large datasets through parallel execution on a cluster of machines. Files are split into smaller pieces to be distributed through the cluster during the \emph{partition step}. Each fragment is distributed to an instance that will execute the \emph{map algorithm} (see Algorithm \ref{alg:map}) and the \emph{reduce algorithm} (see Algorithm \ref{alg:reduce}), sequentially. The output is put forward to the \emph{Planning Module} for rule-based analysis on the consolidated information.

\begin{algorithm}
\caption{Analysis Algorithm: \emph{Map} portion}
\KwData{data packet}
\KwResult{Attack Alert, Data Packet}
\Begin{
Separate packets by protocol into a Hash Map;
\Repeat{end of packages} {
Checks whether the packet has a known attack for this protocol \;
\uIf{there is attack}{Saves IP source, IP Destination, Protocol, Attack \;}
\Else{next package \;}
}
}
\label{alg:map}
\end {algorithm}

\begin{algorithm}
\caption{Analysis Algorithm: \emph{Reduce} portion}
\KwData{Attack Alerts, Data Package}
\KwResult{Source, attack target, protocol, signature, attack type, amount}
\Begin{
Separate packets per protocol in a Hash Map;
\Repeat {end of packages}{
\uIf {Already Registered}{
sum amount of retries}
\Else{adds new attack \;}
}
}
\label{alg:reduce}
\end{algorithm}

\subsection{Response Strategy}
\label{sec:response}

The response strategy follows the concept of \emph{expected utility principle} \cite{briggs2017normative}. Decisions are made based on the probability of positive intrusion events \emph{versus} uncertainty about response effectiveness \cite{bordley2009decision}. For the decision process, the model takes in consideration environmental elements, such as: parameters of the cloud environment; target virtual machines; parameters of the attack set, and; response parameters such as cost of actions, effect time, success probability, effectiveness history, and others. The system model includes the following parameters:

\begin{center}
$\mathbb{N}^* = \{n \in \mathbb{N} | n> 0 \}$: environmental parameters;

$\mathcal{E} = \{e_i | i: states, i \in \mathbb{N}^* \}$: parameters of the attack set;

$\mathcal{A} = \{a_i | i: \textrm{actions}, i \in \mathbb {N}^* \}$: parameters of the response set.

$\mathcal{O} = \{o_i | i: result, i \ in \ mathbb {N} ^ * \}$: result of the actions.
\end{center}

The expected utility formula is given as follows:

\begin{center}
\begin{equation} \label{eq:mathclap_UE}
UE(a_i) = \sum_{o_i \in O} P_{a_i} (o_i).U (o_i)
\end{equation}
\end{center}

where: $\mathcal{O}$ is the result set; $P_{a_i} (o_i)$ is the probability of the result $o_i$ conditioned to the action $a_i$; and $U(o_i)$ is the utility of the result (response) $o_i$. Translating to the construction of the AIRS system, we have the components defined as:

\begin{center}
\noindent $\mathcal{E}$: set of attacks. \\
\noindent $\mathcal{A}$: set of actions that are responses to attacks.\\
\noindent $\mathcal{O}$: result set depending on whether or not they work.\\
\noindent $\mathcal{C}$: set of costs of executing (processing) actions-responses to attacks.\\
\noindent $\mathcal{T}$: set of time durations for action-response execution. \\
\noindent $\mathcal{P}$: set of probabilities of a result to be action-response success. \\
\end{center}

Other definitions involve costs, elapsed times and probabilities, as follows:

\begin{center}
\noindent $\mathcal{C} = \{c_i \ | \ i: cost of action, i \in \mathbb{N}^* \}$
\\
\noindent $\mathcal{T} = \{t_i \ | \ i: time of action, i \in \mathbb{N}^* \}$
\\
\noindent $\mathcal{P} = \{p_i \ | \ i: probability, i \in \mathbb{N}^* \}$
\\
\end{center}

Given these definitions, the \emph{expected utility} $UE(a_i)$ of a response-action $a_i$ considering a result $o_i $ is given by:

\begin{center}
\begin{equation}\label{eq:mathclap_sum}
UE(a_i) = \sum_{i=1}^{m \in \mathbb{N}^*} \frac{p(o_i) }{(c_i + t_i + 1)}
\end{equation}
\end{center}

Where $m$ is the number of actions defined in the system; $a_i \in A$ is an action-response; $p(o_i)$ is the effectiveness probability of $o_i$, and $U(o_i)$ is the expected utility of $o_i$. Normalisation is implemented by shifting the values of each resource so that the minimum value is $0$ and then dividing by the new maximum value, which is the difference between the original maximum and minimum values. The system applies the following method to normalise the utility calculation:

\begin{center}
\begin{equation}\label{eq:minmax}
z = \frac{x - min (x)}{[max(x) - min(x)]}
\end{equation}
\end{center}

Let $\mathcal{A}$ be the set of all possible actions in a processing environment, such that an $a_i \in \mathcal{A}$ element is an action that can be performed, and; $\mathcal{R}$ represents the set of possible responses. Then, a response is defined as a possibly effective action where: $O$ is the result set; $P(a)$ is the conditional probability of the result $o$ given the action $A$; and $U(o)$ is the utility of $o$.

\begin{table}[h]
\label{tab:table-utility}
\centering
\caption{Utility Calculation Example}
\resizebox{\columnwidth}{!}{
\begin{tabular}{|l|l|l|l|l|l|}
\hline
$\mathcal{A}$ & \multicolumn{2}{c|}{State} \\ \cline{2-3} 
      \multicolumn{1}{|c|}{}        & $Attack_{1}$   & $Attack_{2}$  & ...   & $Attack_{n}$   & Expected Utility \\ \hline\hline
$a_1$  &  $UE(a_{11})$  &  $UE(a_{12})$ & ...   & $UE(a_{1n})$   & $UE(a_1) = \sum UE(o_{1n})$  \\ \hline
$a_2$  &  $UE(a_{21})$  &  $UE(a_{22})$ & ...   & $UE(a_{2n})$   & $UE(a_2) = \sum UE(o_{2n}))$  \\ \hline   
$a_3$    & $\frac{p(o_{31}) }{(c_{31} + t_{31} + 1)}$&$\frac{p(o_{32} }{(c_{32} + t_{32} + 1)}$& ...   &  $\frac{p(o_{3n}) }{(c_{3n} + t_{3n} + 1)}$   &  $UE(a_{3})$ = $\sum UE(o_{3n})$  \\ \hline
$a_m$ & $\frac{p(o_{m1}) }{(c_{m1} + t_{m1} + 1)}$ & $\frac{p(o_{m2}) }{(c_{m2} + t_{m2} + 1)}$ & ...   & $\frac{p(o_{mn}) }{(c_{mn} + t_{mn} + 1)}$  &   $UE(a_{m})$ = $\sum UE(o_{mn})$ \\ \hline \hline
\end{tabular}
}
\end{table}

In sum, the method applies cost $c_{in}$, time $t_1n$ and probability $p_in$, where $n$ is the possible number of attacks. The utilities $U_(a_mn)$ is calculated as:

\begin{equation}\label{eq:mathclap_frac}
U(a_{mn}) = \frac{p(a_{mn}) }{(c_{mn} + t_{mn} + 1)}
\end{equation}

Where $m$ corresponds to an action-response; $n$ corresponds to a determined attack as per the examples in Table \ref{tab:table-utility}; $O$ is the result set; $P(a)$ is the conditional probability of the result $o$ given the action $A$; and $U(a)$ is the utility of $a$.

In the proposed model, the largest sum $\max U(a_i)$ corresponds to the \emph{most useful action-response}, which means that we inferred a preferable action-response $a_i$. That is, $a_i$ is more likely to be effective than its peers.


\subsection{Application Example}

Table \ref{tab:examplok} presents an application example of the \emph{expected utility} method considering: a knowledge base $\mathcal{K}$ with previously estimated \emph{utility costs} $\mathcal{C}$, elapsed time $\mathcal{T}$, and success probabilities $\mathcal{P}$, for the attacks $\mathcal{E}$.

\begin{table}[ht]
\centering
\caption{Example of Knowledge Base}
\label{tab:examplok}
\begin{tabular}{|c|c|c|c|c|c|c|c|c|c|}
\hline
\emph{} & \multicolumn{3}{c|}{\emph{$\mathcal{E}_1$}} & \multicolumn{3}{c|}{\emph{$\mathcal{E}_2$}} & \multicolumn{3}{c|}{\emph{$\mathcal{E}_3$}} \\ \hline
\emph{$\mathcal{A}$} & \emph{$\mathcal{T}$} & \emph{$\mathcal{C}$} & \emph{$\mathcal{P}$} & \emph{$\mathcal{T}$} & \emph{$\mathcal{C}$} & \emph{$\mathcal{P}$} & \emph{$\mathcal{T}$} & \emph{$\mathcal{C}$} & \emph{$\mathcal{P}$} \\ \hline
$a_1$  &  2  &   2  &  0,1  &  3  &  4  &  0,3  &  20  &  10  &  0,5 \\ \hline
$a_2$  &  5  &   6  &  0,2  &  3  &  5  &  0,1  &  30  &  10  &  0,4 \\ \hline
$a_3$  &  1  &   1  &   0   &  5  &  7  &  0,2  &  20  &  10  &  0,1  \\ \hline
\end{tabular}
\end{table}

Let us assume that attacks $\mathcal{E}_1$ and $\mathcal{E}_3$ are detected. The first step is to normalise the values of $\mathcal{T}$ and $\mathcal{C}$. Table \ref{tab:exemplokNormalizada} presents the normalised values. Next, we apply the utility formula for each set $a_m$, resulting in the values depicted in Table 
\ref{tab:exemploCalculado}.

\begin{table}[ht]
\centering
\caption{Example of normalised knowledge base}
\label{tab:exemplokNormalizada}
\resizebox{\columnwidth}{!}{
\begin{tabular}{|c|c|c|c|c|c|c|c|c|c|}
\hline
\emph{} & \multicolumn{3}{c|}{\emph{$\mathcal{E}_1$}} & \multicolumn{3}{c|}{\emph{$\mathcal{E}_2$}} & \multicolumn{3}{c|}{\emph{$\mathcal{E}_3$}} \\ \hline
\emph{$\mathcal{A}$} & \emph{$\mathcal{T}$} & \emph{$\mathcal{C}$} & \emph{$\mathcal{P}$} & \emph{$\mathcal{T}$} & \emph{$\mathcal{C}$} & \emph{$\mathcal{P}$} & \emph{$\mathcal{T}$} & \emph{$\mathcal{C}$} & \emph{$\mathcal{P}$} \\ \hline
$a_1$ & 0.034 & 0.111 & 0.100 & 0.068 & 0,333 & 0.300 & 0.655 & 1.000 & 0.500   
\\ \hline
$a_2$ & 0.137 & 0.555 & 0.200 & 0.068 & 0.444 & 0.100 & 1.000 & 1.000 & 0.400               \\ \hline
$a_3$ & 0.000 & 0.000 & 0.000 & 0.137 & 0.666 & 0.200 & 0.655 & 1.000 & 0.000               \\ \hline
\end{tabular}
}
\end{table}

\begin{table}[htb]
\centering
\caption{Example of inferred knowledge base}
\label{tab:exemploCalculado}
\begin{tabular}{|c|c|c|c|}
\hline
& \emph{$\mathcal{E}_1$} & \emph{$\mathcal{E}_2$} & \emph{$\mathcal{E}_3$} \\ \hline
$a_1$ & 0.087 & 0.213 & 0.188 \\ \hline
$a_2$ & 0.118 & 0.066 & 0.133 \\ \hline
$a_3$ & 0.000 & 0.110 & 0.037 \\ \hline
\end{tabular}
\end{table}

Next, you must make the \ref{eq:mathclap_sum} sum of the utilities that correspond to the $a_m$ actions for the $\mathcal{E}_n$ attacks, and choose the highest value utility.

\begin{table}[htb]
\centering
\caption{Utility Calculation}
\label{tab:exemploMax}
\begin{tabular}{|c|c|c|l|}
\hline
 & $\mathcal{E}_1$  &  $\mathcal{E}_3$  &  \emph{$\sum$} \\ \hline
$a_1$ & 0.087 & 0.188 & 0.275 \\ \hline
$a_2$ & 0.118 & 0.133 & 0.251 \\ \hline
$a_3$ & 0.000 & 0.037 & 0.037 \\ \hline
\end{tabular}
\end{table}

In the example above, the action $a_1$ has the best utility and will be selected within the proper context. Moreover, it is possible to adjust the values based on observation of effectiveness of actions in given contexts. One important factor to observe is the elapsed time to implement the actions aiming at effective and efficient delivery. 

%% file: 04results.tex

We developed a proof-of-concept implementation to evaluate the approach and executed it in two scenarios: (i) VMs running on a private cloud in our Lab, and; (ii) VM running on Amazon public cloud. For both cases, we generated two sets of data representing (a) legitimate access and (b) security attacks. The implementation utilises Java 8 and the JnetPCap library. Dedicated attack nodes and legitimate nodes were used to perform the tests. The experiment generated considerable amount of data and demanded extensive processing time. 

The script to generate attacks was implemented upon \emph{Scapy}~\cite{biondi2005packet}, which allows to generate network traffic and inject attacks. The script dynamically mounts a TCP packet informing data, such as source port, destination port, source ip, destination IP, payload, and the \emph{ack} package. The configuration included the target machine and a payload parameter of \emph{QLInject} with large data volumes, forming a typical DDoS attack, presented in Code \ref{lst:script_attack}.

Sensors were installed at specific capturing points to test the \emph{Monitoring Module}. Then, the environment was configured to choose the network interface to be monitored. The captured data was stored in files containing packages. Two criteria were established: elapsed time and data volume. If one of the criteria was triggered, the monitoring files would be sent to the analysis module. The \emph{SARI process} entails: 
\begin{itemize}
    \item collecting the \emph{log} file from the VMs;
    \item transferring the files to the detection server, and;
    \item executing the detection algorithm.
\end{itemize}  

\lstset{language=python,
      basicstyle=\ttfamily\scriptsize,
      keywordstyle=\color{blue}\ttfamily,
      stringstyle=\color{red}\ttfamily,
      commentstyle=\color{green}\ttfamily,
     breaklines=true
     }
 \begin{tcolorbox} 
\begin{lstlisting}[label=lst:script_attack, caption=Script for Attack Simulation,language=python]
from scapy.all import *

seq  = 12345
sport = 1040
dport = 80

ip_packet = IP(dst='172.31.17.62')
syn_packet = TCP(sport=sport,dport=dport,flags='S',seq=seq)

packet = ip_packet/syn_packet
synack_response = sr1(packet)

next_seq = seq + 1
my_ack = synack_response.seq + 1

ack_packet = TCP(sport=sport, dport=dport, flags='A',seq=next_seq, ack=my_ack)

send(ip_packet/ack_packet)

payload_packet = TCP(sport=sport, dport=dport, flags='A', seq=next_seq, ack=my_ack)
payload = "GET / HTTP/1.0\r\nHOST: 172.31.17.62\r\n\rSQL INJECT\n"

for i in range(100):
 	sr(ip_packet/payload_packet/payload, multi=1, timeout=1)

\end{lstlisting}
\end{tcolorbox}

\begin{table}[!h]
\centering
\caption{Experimentation datasets}
\label{tab:attacks}
\begin{tabular}{|l|l|l|}
\hline
\emph{\begin{tabular}[c]{@{}l@{}}Size in MB\end{tabular}} & \emph{\begin{tabular}[c]{@{}l@{}}Number \\ of Packets \end{tabular}} & \emph{\begin{tabular}[c]{@{}l@{}}Number \\ of Attacks \end{tabular}} \\ \hline
10 & 130,000 & 306
\\ \hline
50 & 700,000 & 803
\\ \hline
100 & 1,400,000  & 1,481
\\ \hline
200 & 3,625,000 & 1,851
\\ \hline
400 & 7,250,000 & 2,138
\\ \hline
600 & 8,845,000 & 2,822 
\\ \hline
800 & 11,600,000 & 3.442
\\ \hline
1.000 & 14,500,000 & 4,275
\\ \hline
\end{tabular}
\end{table}

Table \ref{tab:attacks} presents variations of datasets generated through multiple experimentation configurations.

The \emph{Analysis and Planning} module is implemented in Java using the \emph{Hadoop} library to support MapReduce. For each experiment a \emph{cluster} was created to process the analysis. This module receives the datasets for processing and generates results like the one depicted in Table \ref{tab:resultadoAnalise}. Table 
\ref{tab:paramutilidade} presents the parameters for the utility calculation.

\begin{table}[!h]
\centering
\caption{Result of the Analysis Module}
\label{tab:resultadoAnalise}
\resizebox{\columnwidth}{!}{
\begin{tabular}{|l|l|l|l|l|}
\hline
 & \emph{Source IP} & \emph{IP Destination} & \emph{Attack} & \emph{Quantity} \\ \hline
1 & 150.162.63.200 & 150.162.63.23 & $\mathcal{E}_1$ & 275 \\ \hline
2 & 150.162.63.205 & 150.162.63.23 & $\mathcal{E}_2$ & 2000 \\ \hline
3 & 150.162.63.202 & 150.162.63.23 & $\mathcal{E}_3$ & 2000 \\ \hline
\end{tabular}
}
\end{table}

\begin{table}[!h]
\centering
\caption{Parameters for the Utility Calculation}
\label{tab:paramutilidade}
\resizebox{\columnwidth}{!}{
\begin{tabular}{|l|l|l|l|}
\hline
\emph{Actions} & \emph{Probability} & \emph{\begin{tabular}[c]{@{}l@{}} \\ Normalised Cost \end{tabular}} & \emph{\begin{tabular}[c]{@{}l@{}} \\ Normalised Elapsed Time \end{tabular}} \\ \hline
$a_1$ & 0.01 & 0.00157 & 0.01023 \\ \hline
$a_2$ & 0.15 & 0.00166 & 0.01063 \\ \hline
$a_3$ & 0.34 & 0.00375 & 0.02127 \\ \hline
$a_4$ & 0.41 & 0.00583 & 0.03191 \\ \hline
$a_5$ & 0.48 & 0.00791 & 0.04255 \\ \hline
\end{tabular}
}
\end{table}

\subsection{Experimenting on a Private Cloud}

\begin{figure}
\centering
\includegraphics[width=6.5cm]{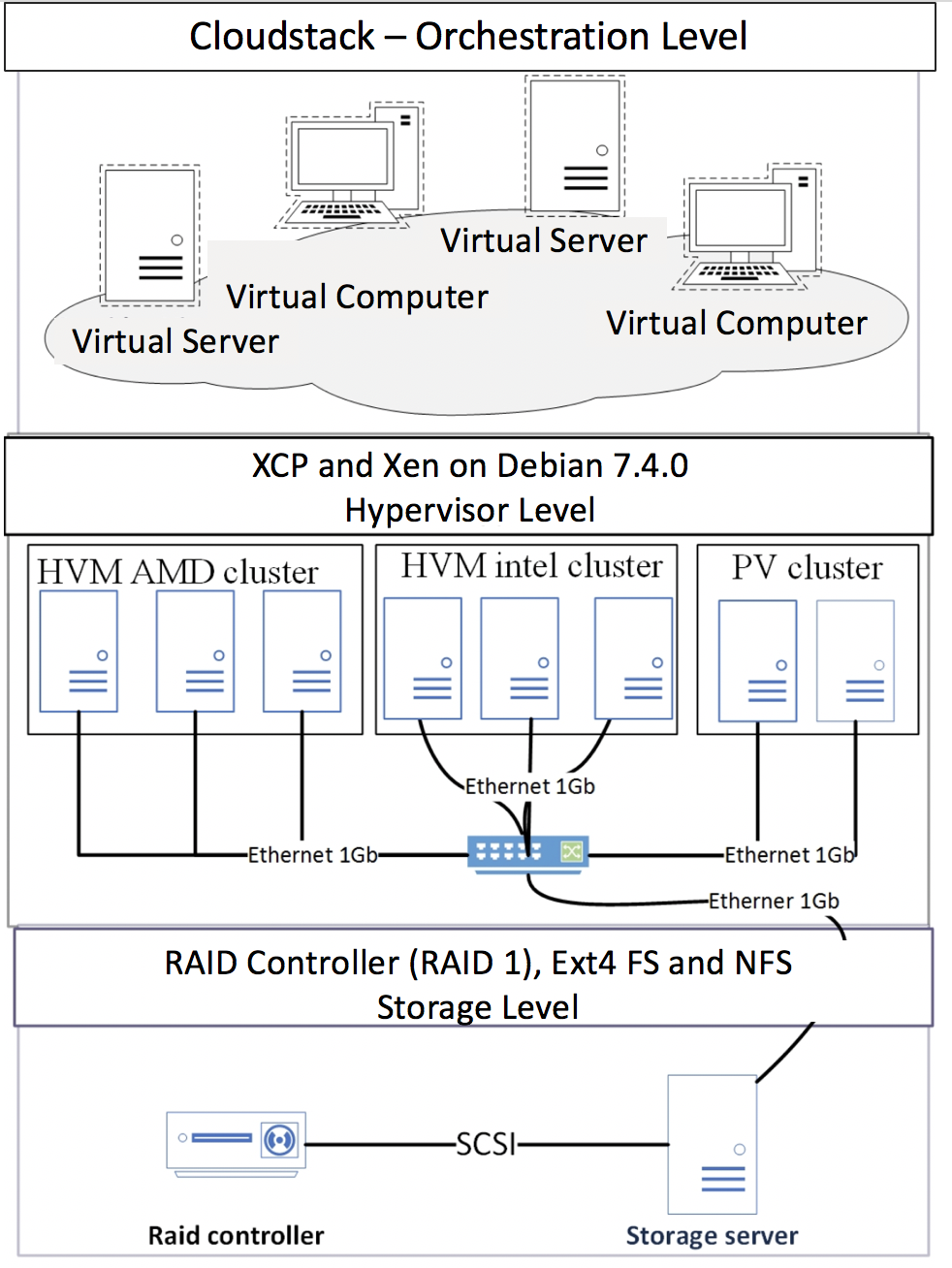}
\caption{Private cloud environment}
\label{fig:ambiente-validacao-lrg}
\end{figure}
 
Figure \ref{fig:ambiente-validacao-lrg} depicts the testing environment implemented over a private cloud computing in our laboratory. The environment is composed of a \emph{CloudStack} hypervisor and \emph{Xen} orchestration system running on \emph{Debian}. We created: (a) a set of VMs representing the invaders; (b) a set of VMs with WEB servers and databases representing the target, and; (c) a cluster of 3 computers to execute SARI processes. 

\begin{figure}
\centering
\includegraphics[width=\linewidth]{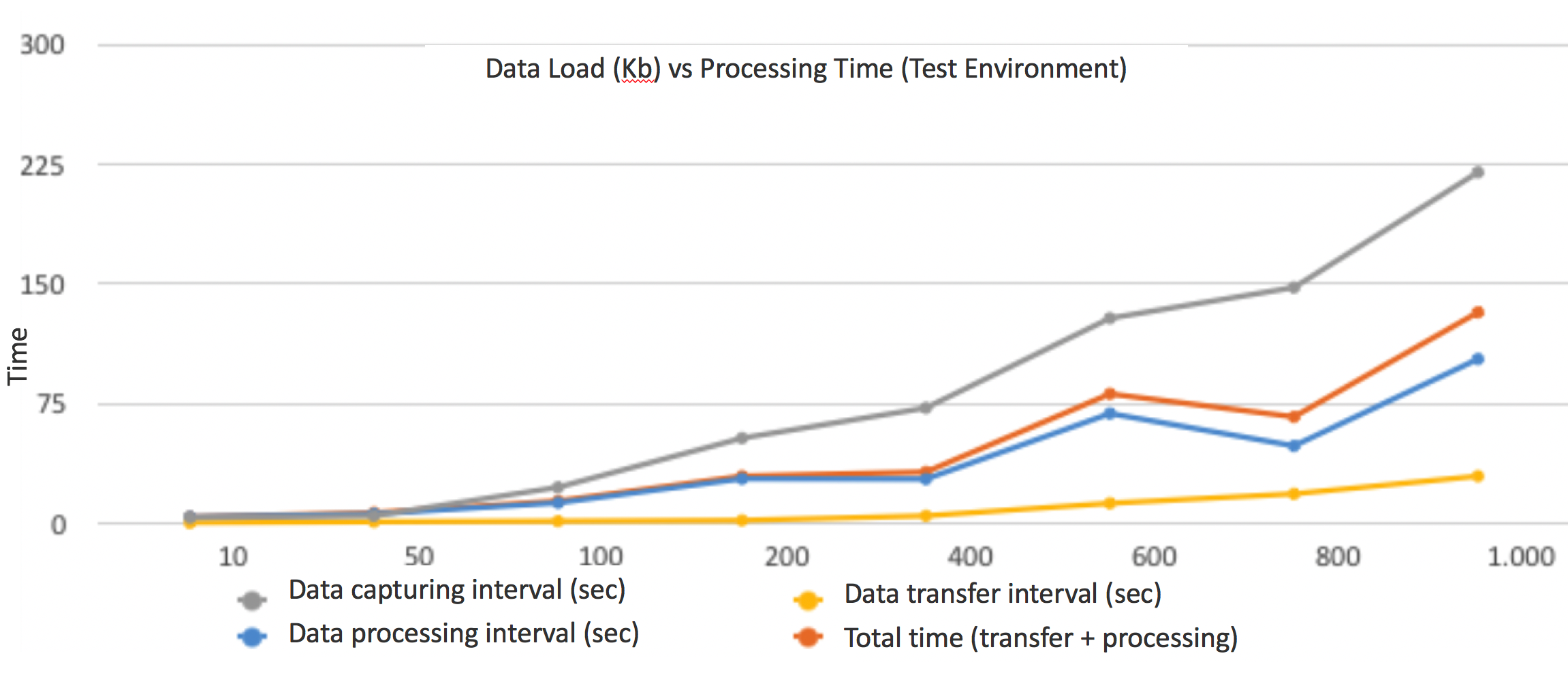}
\caption{Total processing time on the Private Cloud}
\label{fig:totalLRG}
\end{figure}
 
\begin{figure}
\centering
\includegraphics[width=\linewidth]{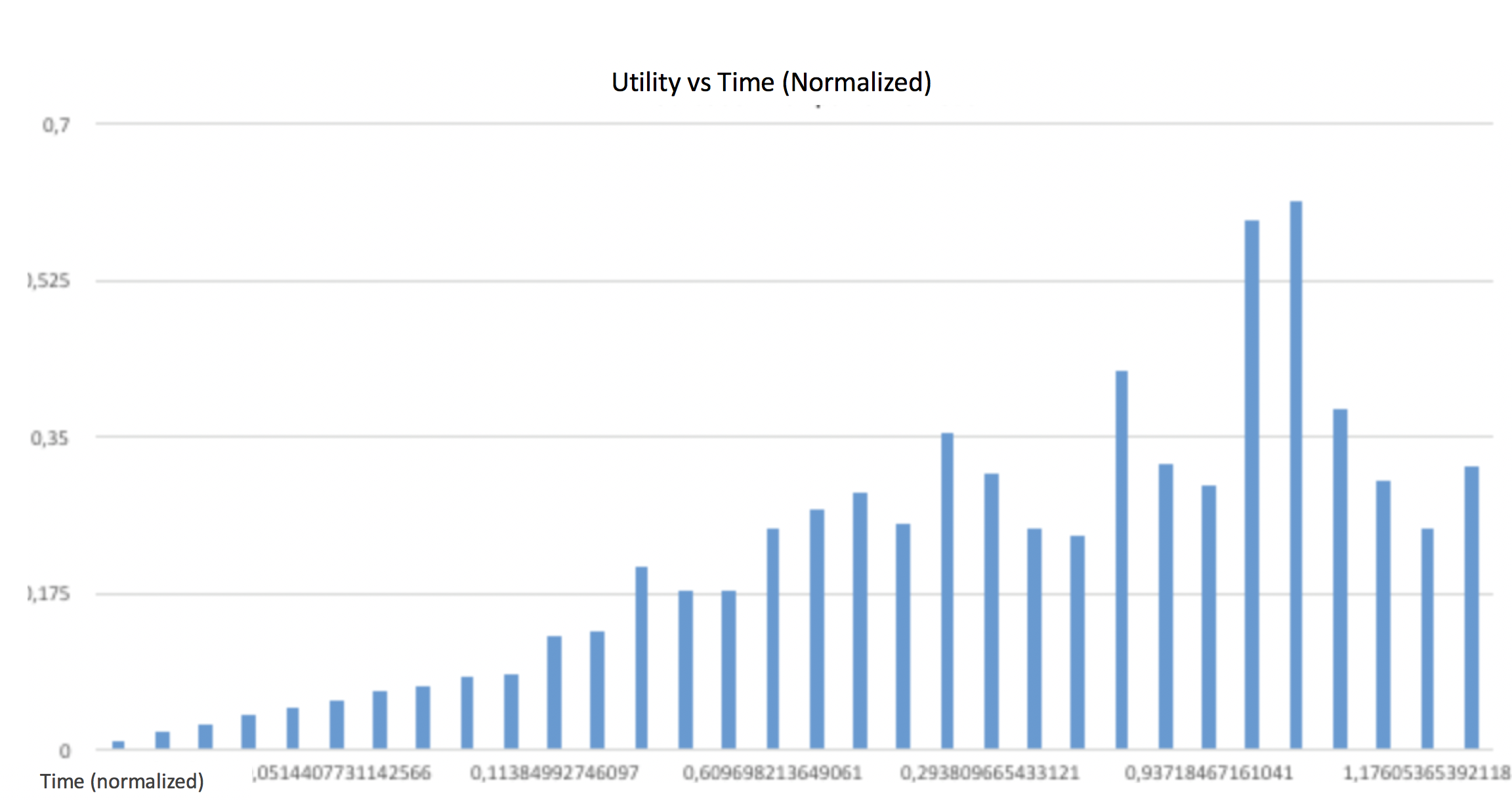}
\caption{Ratio utility \emph{versus} cost on the Private Cloud}
\label{fig:utilityLRG}
\end{figure}
 
Figure \ref{fig:totalLRG} shows the behaviour of the analysis module against the different configurations. Figure \ref{fig:utilityLRG} depicts the normalised ratios in the testing environment experiments. There are two values close to the maximum utility (0.637 and 0.650), since they represent actions where the probability is the same (95 \%). The execution the cost presents small variation (0.443 and 0.451). However, the execution time presents ample variation (9.118 and 42.011). We conclude that the thread with the smallest processing time has the greatest utility (0.650).

\subsection{Experimenting on a Public Cloud}

The public cloud experimentation setup is similar to the private cloud one. The implementation was on the Amazon Web Services (AWS) platform, simulating legitimate and intrusive users against 4 target machines as service providers.The SARI system is represented by an interface for monitoring and analysis, along with a cluster for planning and execution and the knowledge base.

\begin{figure}
\centering
\includegraphics[width=\linewidth]{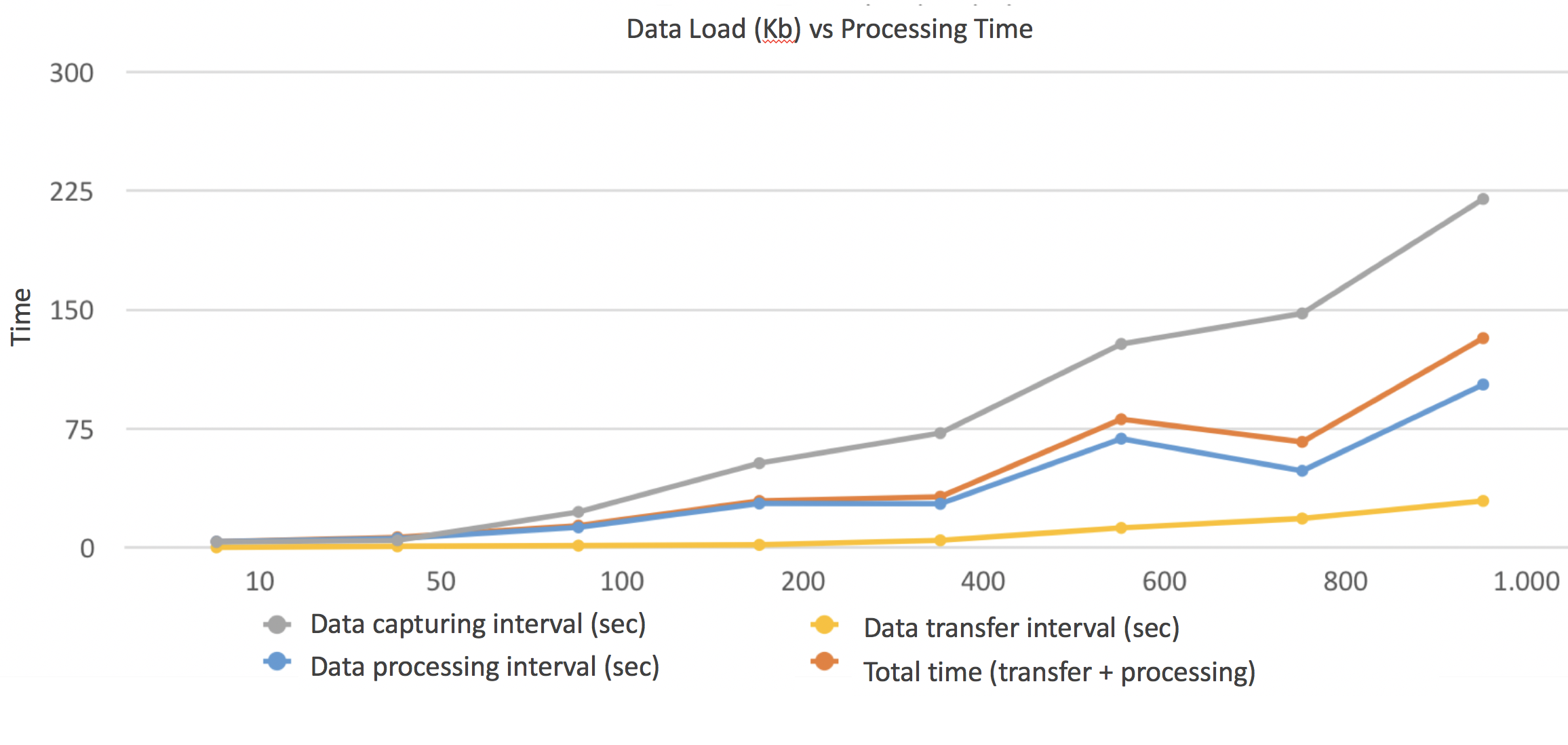}
\caption{Total processing time on the Public Cloud}
\label{fig:totalAWS}
\end{figure}

\begin{figure}
\centering
\includegraphics[width=\linewidth]{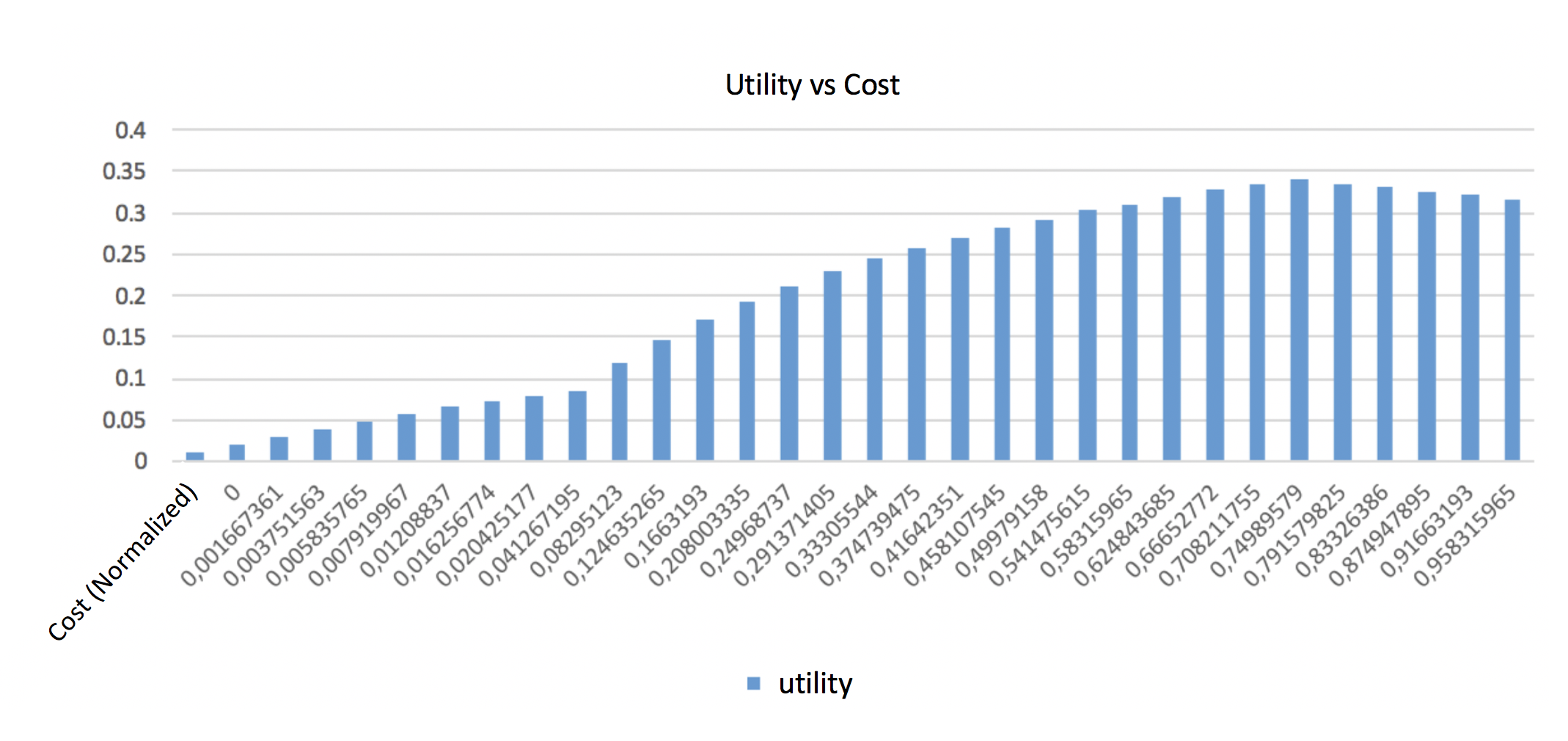}
\caption{Ratio utility \emph{versus} cost on the Production Cloud}
\label{fig:utilityAWS}
\end{figure}

Figure \ref{fig:totalAWS} presents the results from the execution on the public cloud. Figure \ref{fig:utilityAWS} presents the utility function for the cost of actions. The most useful action has value $U = 0.340309094$ and was process in 950 seconds at a cost of 1900 units. 

The key difference between the public and private cloud experiments is the time lag between the steps, which is shorter in the public cloud environment due to the larger availability of computational resources. 

Hence, we concluded for significant improvement in response effectiveness and potential to scale to large environments.

%% file: 05conclusion.tex

We presented a reference architecture for Automated Intrusion Detection based on methods of \emph{Big Data} for the  classification, understanding and prediction of behavioural deviance in Distributed Computing environments. The proposed solution covers for the technology gap attack detection strategies that provide satisfactory results in Distributed Computing environments.

The \emph{Autonomic Intrusion Response System} follows the vision of \emph{autonomic computing} and works based on the Monitor-Analyse-Plan-Execute-Knowledge (MAP-K) architecture to efficiently analyse large amounts of data about the utilisation of Distribute Computing resources. The solution employs a knowledge based approach to detect known attacks by comparing attack signatures to suspicious actions The strategy applies \emph{MapReduce} to allow working on large datasets through parallel execution on a cluster of machines. 

We evaluated the proposed approach through a prototype implementation against two scenarios: (i) VMs running on a private cloud in our Lab, and; (ii) VM running on Amazon public cloud.  The results demonstrate the effectiveness of the solution allowing to process large volumes of access information in acceptable time delays for both the private cloud and the public cloud experiment. We demonstrated that the approach is able to handle real-world scenarios and deliver low latency response results, aligned with the requirements for Automated Intrusion Detection. Hence, we concluded for significant improvement in response effectiveness and potential to scale to large environments.

We argue that a product-grade implementation based on the proposed reference architecture would effectively reduce the damage caused by diverse forms of Cyber attacks on Distributed Computing environments. 

As a limitation, the proposed approach does not contemplate optimisation upon the algorithmic complexity of the expected utility theory. That is, given an attack, the algorithm needs to calculate the sum of the utility of each response. This calculation grows exponentially given the number of responses implemented in the model. Another limitation is the application of rules for knowledge-based detection methods using known attacks within the scope of this work. We consider these limitations acceptable, as the purpose of the project was to demonstrate the feasibility of using Big Data strategies and provide a reference architecture for the implementation. Further work will have to extend on this discussion to attain product-grade implementations.

Further work may also involve research in the application of Machine Learning and Cognitive Computing to detect attacks beyond the scope of the implemented rules. New \emph{attack signatures} could be discovered and incorporated to the architecture's knowledge base thus continuously improving the system's effectiveness overtime. This strategy would progress the solution towards a self-learning and self-adjustable system, laying the ground for a future \emph{Cognitive Intrusion Detection Systems}.

%% file: 06acknowledge.tex

This work was conducted by Dr. Kleber Vieira in the scope of his doctorate program in Computer Sciences at the Network and Management Laboratory (LRG), Department of Informatics and Statistics, Federal University of Santa Catarina (UFSC), Brazil. The research was supervised by Prof. Dr. Carlos Becker Westphall and counted with valuable input from LRG's colleagues. Special thanks to Prof. Dr. Joao Bosco Sobral and Prof. Dr. Jorge Lopes de Souza Leao for their input and contribution. Dr. Fernando Koch contributed with putting this paper together and provided extensive input during the elaboration of the research. Dr. Koch is a Visiting Researcher at LRG/UFSC, Honorary Senior Fellow with The University of Melbourne, Australia, and supported by the Brazilian CNPq Productivity in Technology and Innovation Grant (CNPq 307275/2015-9).